\documentclass[10pt,conference]{IEEEtran}

\usepackage{color}
\usepackage{verbatim}
\usepackage{afterpage}
\usepackage{url}
\usepackage{xfrac}
\usepackage{algorithm}
\usepackage{algorithmicx}
\usepackage{algpseudocode}
\usepackage{amsmath}
\usepackage{amsfonts}
\usepackage{amssymb}
\usepackage{braket}
\usepackage{graphicx}
\usepackage{braket}
\usepackage[nocompress]{cite}
\usepackage{booktabs}
\usepackage{multirow}
\usepackage{caption}
\usepackage{subcaption}
\usepackage{cite}

\long\def\comment#1{}

\def\parah#1{\vspace*{0.0in} \noindent{\bf #1:}}

\IEEEoverridecommandlockouts

\DeclareMathOperator*{\argmax}{argmax}
\newcommand{\iswap}{$\sqrt{\text{iSWAP}}$}

\begin{document}

\title{Quantum Circuit Optimization and Transpilation via Parameterized Circuit Instantiation}

\author{
	\IEEEauthorblockN{Ed Younis, Costin Iancu}
	\IEEEauthorblockA{
	    \{edyounis,cciancu\}@lbl.gov\\
		Lawrence Berkeley National Laboratory\\
	}
}

\maketitle

\begin{abstract}
Parameterized circuit instantiation is a common technique encountered
in the generation of circuits for a large class of hybrid
quantum-classical algorithms. Despite being supported by  popular
quantum compilation infrastructures such as IBM Qiskit and Google
Cirq, instantiation has not been extensively considered in the context of
circuit compilation and optimization pipelines. In this work, we
describe algorithms to apply instantiation during two common compilation
steps: circuit optimization and gate-set transpilation. When placed in
a compilation workflow, our circuit optimization algorithm produces
circuits with  an average of 13\% fewer gates than other optimizing
compilers. Our gate-set transpilation algorithm can target any
gate-set, even sets with multiple two-qubit gates, and produces
circuits with an average of 12\% fewer two-qubit gates than other
compilers. Overall, we show how instantiation can be incorporated into
a compiler workflow to improve circuit quality and enhance
portability, all while maintaining a reasonably low compile time
overhead.
\end{abstract}

% Sections
\section{Introduction}

Circuit instantiation is an operation performed in many
hybrid-classical quantum algorithmic workflows such as VQE~\cite{vqe} or
QAOA~\cite{qaoa}. In this setting, the algorithm has associated a 
circuit ansatz, where gates are represented in their
parameterized form, e.g. $R_x(\theta)$. The execution iterates, and at
each step the gate parameters are instantiated (often by a numerical
optimizer) using results from previous quantum executions. Due to
the popularity of hybrid algorithms, circuit instantiation is directly
supported by popular compilation infrastructures such as IBM Qiskit~\cite{qiskit},
Google Cirq~\cite{cirq} or Tket~\cite{tket}.

Instantiation is also becoming an important step in circuit synthesis
tools such as BQSKIT~\cite{bqskit}(QSearch~\cite{qsearch} and QFAST~\cite{qfast}), NACL~\cite{nacl}, and Squander~\cite{squander1, squander2}.
Here, circuits are built bottom-up using parameterized gates
and are instantiated using numerical optimization at each step.

In this paper, we demonstrate that instantiation is a  powerful
primitive that can be easily incorporated into any quantum compilation
workflow. In our approach, given an input circuit, we replace its
gates with parameterized representations and apply
instantiation to improve the resulting quality for two essential tasks: circuit optimization and gate-set transpilation.

\parah{Circuit Optimization} Circuit depth is a direct measure of quantum program performance. Errors compound its impact in the NISQ era, and as a result, a significant portion of the compilation workflow is dedicated to circuit depth reduction. Our algorithms reduce circuit depth by iteratively deleting gates and reinstantiating the circuit until convergence criteria are met. As each gate in a circuit becomes parameterized, numerical optimizers face scalability problems for large circuits. For these, we use a circuit partitioning approach introduced in~\cite{qgo} combined with iterative deletion.

\parah{Gate-set Transpilation} Any available quantum processor architecture exposes to users a reduced native gate-set: a few single-qubit rotations and usually a single two-qubit gate. Programs are often generated or expressed in only one gate-set, e.g., \{$R_z$, SX, CNOT\} for some IBM processors. When transpiling a program to a different native gate-set, most compilers convert two-qubit gates using fixed gate identities or KAK~\cite{kak} decompositions. In some cases, translations from one gate to another are not readily available. Indeed, when considering the set of existing native two-qubit gates \{CNOT, CZ, XX, ZZ, \iswap, SYC\}, we note that the most widely used
compilers have limited support for transpiling between these gates. Our instantiation-based formulation is
completely portable as we directly represent gates by their
associated unitary matrix.

The main contributions of this work are as follows:
\begin{enumerate}
    \item A formal definition of instantiation and a brief survey of it in quantum compiler literature.
    
    \item An optimization algorithm that performs well in existing compilation pipelines and introduces non-local optimizations not present in current passes.
    
    \item A quantum circuit retargeting algorithm that works well on any gate-set without user-defined rules. The algorithm can be further specialized with user rules, if so desired. It even works on gate-sets with multiple two-qubit gates and produces circuits shorter than every other algorithm. The transpiled circuits are informative about the efficacy of a particular gate-set to represent algorithms.
    
    \item A verification procedure for non-exact compilation algorithms on very large circuits that are not simulatable.
  \end{enumerate}

  We evaluate our approach on a set of circuits used by other researchers to assess the efficacy of compiler optimizations~\cite{supermarq,tket,qiskit}. For comparison, we consider the Qiskit, Cirq, and Tket compilers.
  When adding our instantiation-based optimization to already existing optimizing workflows, we observe an additional gate reduction of 5\% for two-qubit gates and 23\% for single-qubit gates. When transpiling to existing hardware gate-sets, we observe an average final two-qubit gate reduction of 12\%. In particular, when we were transpiling a 64-qubit time-evolution circuit to Google's Sycamore gate (SYC), we recorded a circuit with 3970 two-qubit gates, which is a 51\% reduction over Cirq's circuit with 8064 two-qubit gates.

When examining the execution time overhead, we observe an average slowdown of $14\times$; however, we demonstrate the tunability of our algorithms by speeding up a specific execution by $13\times$ while only reducing quality by 0.6\%.

The following intuition provides an explanation for the quality of our results. During circuit optimization, the available compilers use a sequence of rule-based peephole transformations, where at each step, the circuit is only transformed locally. Similarly, for transpilation, they form two-qubit blocks and either apply rule-based translations or KAK-based decompositions. In contrast, a circuit undergoes many global transformations with our approach.

Overall, we believe that the results indicate that instantiation can be easily and safely incorporated as a step in the compilation workflow. The ability to transpile algorithms well between native gate sets enables interesting architectural comparisons. For example, 
most circuits transpiled to \iswap\  and Sycamore gates required more gates than in other gate-sets. On the other hand, due to the low gate latency~~\cite{google_hardware}, these long circuits may still yield faster execution and potentially less error.

The rest of this paper is structured as follows. In Section~\ref{sec:inst}, we introduce parameterized circuit instantiation and provide a brief survey of it in compiler literature. Then we describe our optimization and retargeting algorithms in Section~\ref{sec:alg}. We include our experimental setup and verification procedure in Section~\ref{sec:exp} and evaluate the algorithms in Section~\ref{sec:eval}. Lastly, we discuss the results in Section~\ref{sec:disc} and conclude in Section~\ref{sec:conc}.
\section{Parameterized Circuit Instantiation}
\label{sec:inst}

\begin{figure}
    \centering
    \includegraphics[scale=1.5]{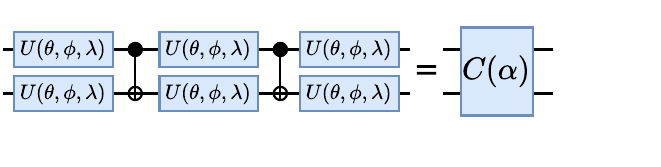}
    \caption{\footnotesize \it This is an example of a parameterized quantum circuit on the left. It is composed of three-parameter universal single-qubit rotations and two-qubit CNOT gates. For simulatable circuits, we can represent the circuit by its unitary operator shown on the right, which is calculated by tensor contraction of all of its gates. Furthermore, we can represent parameterized circuits by a parameterized unitary $C(\alpha)$, which can be instantiated to some other unitary $V$ by solving for the parameters $\alpha$ that maximize the overlap of $C(\alpha)$ and $V$. This can be accomplished with analytic methods in specific cases and gradient descent or other numerical methods in the general case. The text describing a parameterized single-qubit rotation will be left out in the other figures; a box on a single wire depicts a generic parameterized single-qubit gate.}
    \label{fig:params}
\end{figure}

Instantiation is the process of finding the parameters for
a circuit's gates that make it to most closely implement
a target unitary. Techniques that perform instantiation are ubiquitously
deployed in quantum compiler toolchains. The formal problem definition is given by:

\parah{Parameterized Circuit Instantiation Problem}
Given a parameterized quantum circuit $C: \mathbb{R}^k \mapsto U(N)$ and a target
unitary $V \in U(N)$, solve for $$\argmax_{\alpha}{tr(V^{\dagger}C(\alpha))}$$
where $k$ is the number of gate parameters in the circuit, and $U(N)$ is the set of all $N \times N$ unitary matrices. This definition is very general and considers the parameterized circuit as a parameterized unitary operator, see Figure~\ref{fig:params}. The $tr(V^\dagger C(\alpha))$ component measures the Hilbert-Schmidt inner product, which physically represents the overlap between the target unitary and the circuit's operator. The maximum value this can have is equal to $N$ the dimension of the matrix, and this occurs when $C(\alpha)$, the unitary of the circuit with gate parameters $\alpha$, is equivalent to $V$ the target unitary up to a global phase.

The most common form of instantiation is the KAK~\cite{kak} decomposition, which uses analytic methods to produce the two-qubit circuit that implements  any two-qubit unitary. Compilers have used this decomposition to optimize long sequences of operations. This is done by first grouping together consecutive gates on a pair of qubits, then calculating the unitary implemented by the grouped gates, and finally applying the KAK decomposition to convert to a potentially shorter sequence of gates.

For every universal gate-set, the KAK decomposition can yield a parameterized circuit, to which it can instantiate any two-qubit unitary. Therefore, applying the KAK decomposition to retarget a circuit's gate-set is also possible. Once a template is discovered in the desired gate-set, it can be utilized similarly to the optimization procedure to convert grouped gates to gates of a different type. However, producing a circuit template when designing a new gate-set may be nontrivial.

Recently, bottom-up approaches to quantum synthesis have been successful through numerical instantiation~\cite{qsearch, qfast, nacl, squander1, squander2, bestapprox}.
Rather than fixed mathematical identities, these techniques employ a numerical optimizer to closely approximate a solution to the instantiation problem. This is done by minimizing a cost function, often the unitary error or distance between the circuit's unitary and a target unitary. This is given by the following formula using the same notation as before.

$$\Delta(C(\alpha), V) = 1 - \frac{|tr(V^\dagger C(\alpha))|}{N}$$

Other variations of this distance function include:

$$\Delta_f(C(\alpha), V) = 1 - \frac{Re(tr(V^\dagger C(\alpha)))}{N}$$

and

$$\Delta_p(C(\alpha), V) = \sqrt{1 - \frac{|tr(V^\dagger C(\alpha))|^2}{N^2}}$$

All three methods have a range of $[0, 1]$, and as they approach zero, the circuit's
unitary approaches V. For the rest of our paper, we refer to the unitary
distance or error as $\Delta$, the first formulation.

Bottom-up synthesis methods build up the parameterized circuit by iteratively placing more gates. At every step, they instantiate the parameters using numerical optimizers and claim a solution when the error is less than some $\epsilon$ threshold. For example, QSearch~\cite{qsearch} phrases the synthesis problem as a search over circuit templates, instantiating candidates as it searches over them. QFAST~\cite{qfast} expands on this by conflating the search and numerical optimization using a unique circuit encoding. In the NACL~\cite{nacl} algorithm, machine learning methods are used alongside instantiation to synthesize noise-aware circuits. SQUANDER~\cite{squander2} takes a more direct approach to bottom-up synthesis. Here the circuit template is grown and reinstantiated iteratively until a solution is found, at which point, the circuit is then compressed into its optimal form. In~\cite{bestapprox}, instantiation is used in a similar bottom-up approach to produce approximate circuits.

Several compilation algorithms have leveraged the high quality of results produced during bottom-up synthesis. QGo~\cite{qgo} combined circuit partitioning with synthesis to create a circuit mapping and optimization procedure. QUEST~\cite{quest} uses similar techniques to produce robust and resource-efficient approximate circuits. The growth of algorithms using numerical instantiation has been a motivation for BQSKit~\cite{bqskit} developers to build a Python framework around instantiation.

\section{Numerical Instantiation-based Algorithms}
\label{sec:alg}

\begin{figure*}
    \centering
    
    \begin{subfigure}[t]{0.45\textwidth}
        \includegraphics[width=2.5in,height=2in]{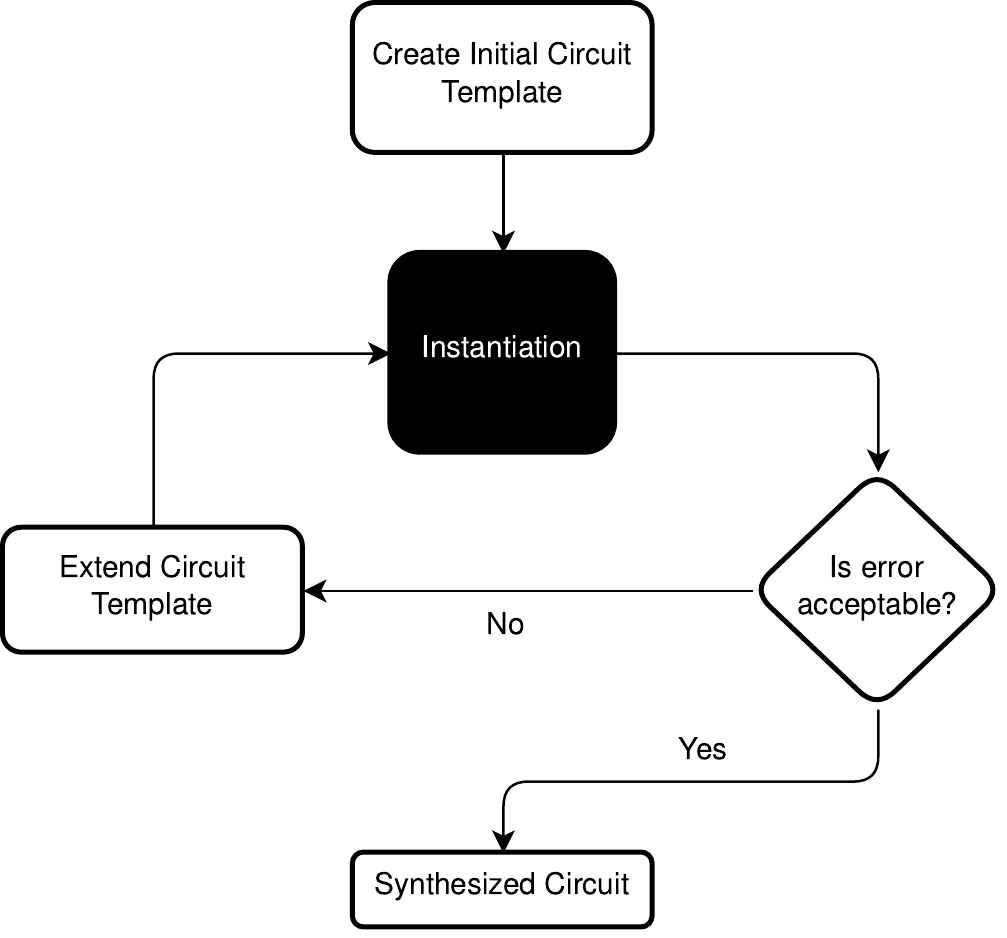}
        \caption{Standard Bottom-up Synthesis Flow}
        \label{fig:bottomup}
    \end{subfigure}
    ~
    \begin{subfigure}[t]{0.45\textwidth}
        \includegraphics[width=2.5in,height=2in]{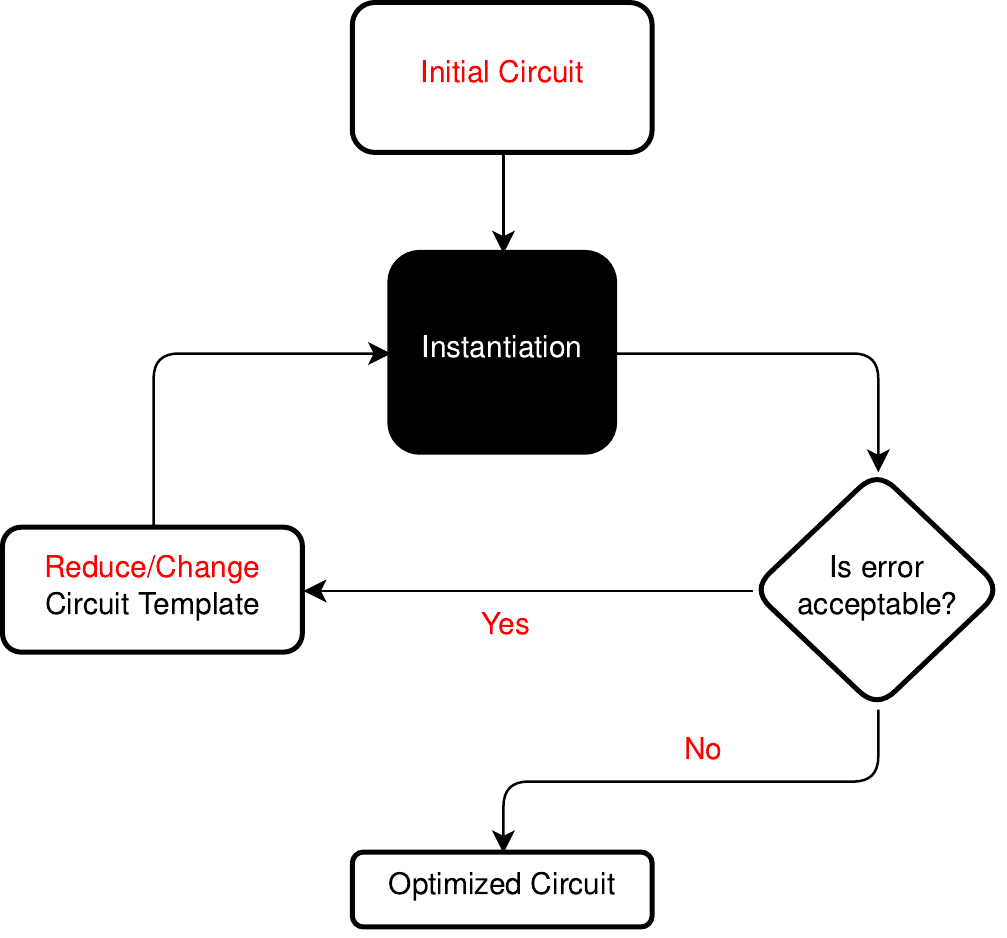}
        \caption{Circuit Optimization or Transpilation Flow}
        \label{fig:optloop}
    \end{subfigure}
    
    \caption{\footnotesize \it
        Instantiation is the core subroutine in all bottom-up circuit synthesis algorithms.
        The standard bottom-up loop (left) is to create an initial circuit
        template guess then continue to instantiate and extend it until an acceptable solution is found.
        Through a slight modification to this flow, instantiation can also be used to transform or optimize a quantum circuit (right).
        Instead of extending a circuit template like in bottom-up synthesis, we reduce
        or transform a circuit template and continue until it is no longer successful or necessary.
    }
    \label{fig:loops}
\end{figure*}

Numerical instantiation is a powerful tool for restructuring circuits
in non-intuitive ways leading to circuit optimizations  likely unattainable by
rule-based applications. Many synthesis algorithms have utilized it
for high-quality circuit construction through some variation of the
bottom-up loop described in Figure~\ref{fig:bottomup}.
We have found small changes to this loop can yield a strategy for
modifying already existing circuits, see Figure~\ref{fig:optloop}.
We demonstrate this strategy by building two algorithms, a {\it circuit
optimization} algorithm and a {\it gate-set retargeting} algorithm.

\subsection{Gate Deletion}

\begin{figure}
    \centering
    \includegraphics[width=\columnwidth]{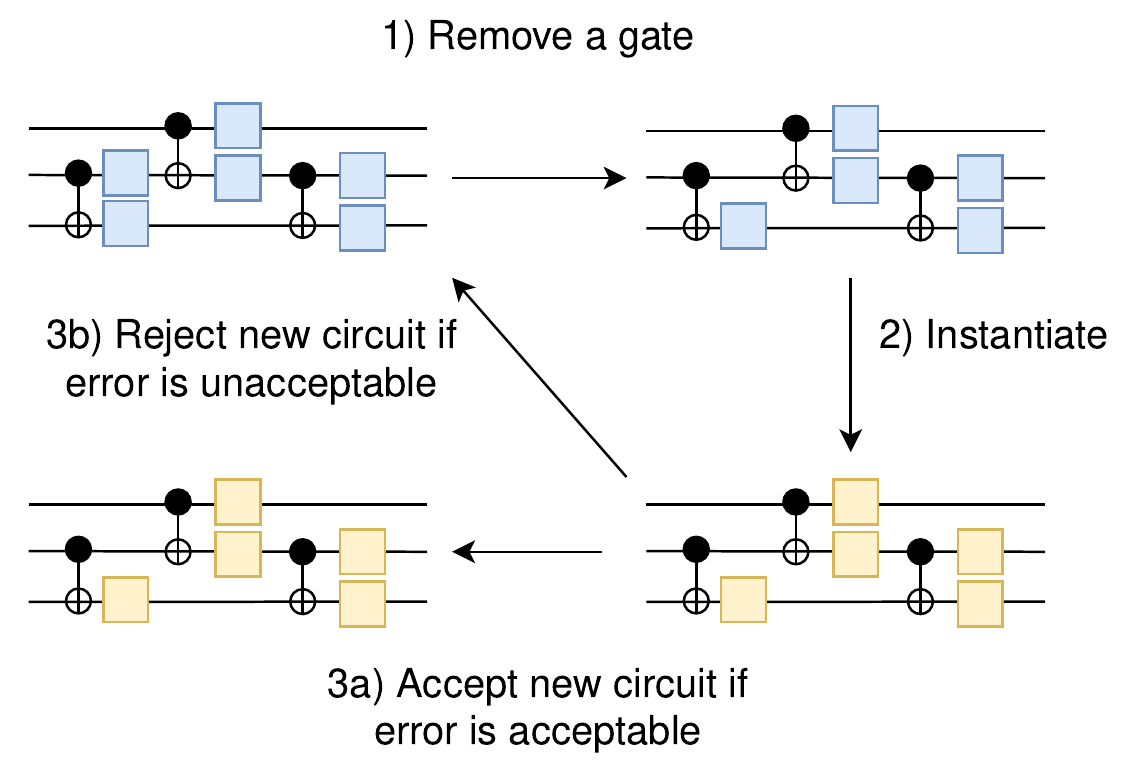}
    \caption{\footnotesize \it
        Gates can be removed from circuits through the use of instantiation.
        This can be done by first selecting and removing a gate, then instantiating
        the remaining gates' parameters to make up for the loss. This is not always
        successful. If the error on the new circuit is less than some threshold then
        the new circuit is accepted, otherwise the circuit is rejected.
    }
    \label{fig:gate_del}
\end{figure}

    The most straightforward way to use the transformation loop described in
    Figure~\ref{fig:optloop} is to remove gates from a circuit.
    This gate deletion process can be done by selecting a gate or
    group of gates to remove and then instantiating the parameters to
    the remaining gates to accommodate the loss in accuracy. See Figure~\ref{fig:gate_del} for an illustration.
    
    It is not always guaranteed that the result from deleting a gate will be an acceptable solution
    to the target unitary. This is partially because not all circuits can implement all unitaries.
    It may be the case that removing a gate ensures the remaining gates can never implement the original target,
    and therefore, even an ideal instantiation tool will never find an acceptable solution.
    Gate deletion can also be fallible because the circuit space is extremely high dimensional,
    and the optimizers can get stuck in local minima. To combat this,
    one can use multi-start optimization~\cite{APOSMM}  strategies during instantiation.
    
\subsection{Optimization Algorithm}

    We need a methodology to select candidate gates for removal to use gate deletion in an optimization algorithm. For the same reasons mentioned previously, different strategies will lead to different results. Additionally, an algorithm's runtime is directly related to the number of times instantiation is performed.
    
    At one end of the spectrum, we have exhaustive search procedures
    that scale exponentially with the number of gates in the starting
    circuit: the method will find the optimal
    result. This can be accomplished by first attempting to remove
    each gate individually and then repeating this step on all
    circuits that successfully removed a gate.
    
    At the other end of the spectrum, a linear complexity method will
    consider deleting each gate only once. A gate that failed to be
    removed will never be reconsidered. Assuming that we have an ideal
    oracle instantiator, this approach will also find the optimal
    result in many cases. However, in practice we often find that
    order of deletion matters: after removing a gate, re-evaluating
    the removal of a previously considered gate can lead to a
    successful deletion. Two observations explain this behavior: 1)
    the instantiation problem is challenging to  existing numerical
    optimizers; and 2) the newer problem is different and it has lower dimensionality, leading to fewer local minima.
    
Our deletion algorithm uses a middle-ground approach where we
iteratively scan the circuit using a linear strategy: in each scan we
attempt to delete each gate only once. The algorithm stops  when no more
gates can be deleted subject to the $\Delta$ accuracy convergence
criteria requested by the user.  Algorithm~\ref{alg:opt_code} provides pseudocode for
    this method.
    
    This algorithm can work on any circuit with any gate-set because it only relies on numerical instantiation and not fixed rules. It can also work on multi-level systems such as qutrits. Furthermore, gate deletion will only ever remove gates, which implies it can be run on a routed circuit without the worry of ruining the routing.
    
\begin{algorithm}
\caption{Iterative Scanning Gate Removal}
\label{alg:opt_code}
\begin{algorithmic}[1]
\State $C \gets \text{input circuit}$
\State $\epsilon_t \gets \text{acceptable error threshold}$
\State $U_t \gets \text{the unitary implemented by }C$
\State $N \gets \text{the number of gates in }C$
\State $M \gets 0$
\While{$N\neq M$} 
    \For{$g \in C$}
        \State Remove $g$ from $C$
        \State $C, \epsilon \gets \text{instantiate}(C, U_t)$
        \If{$\epsilon > \epsilon_t$}
            \State Place $g$ back into $C$
        \EndIf
    \EndFor
    \State $M \gets N$
    \State $N \gets \text{the number of gates in }C$
\EndWhile
\end{algorithmic}
\end{algorithm}

\subsection{Gate-Set Transpilation}

    Another use of the transformation loop described in Figure~\ref{fig:optloop} is to substitute gates with different ones. Instead of deleting a gate, we can remove and replace a gate or group of gates with another gate or group of gates. 
    
\begin{figure*}
    \centering
    \includegraphics[scale=0.9]{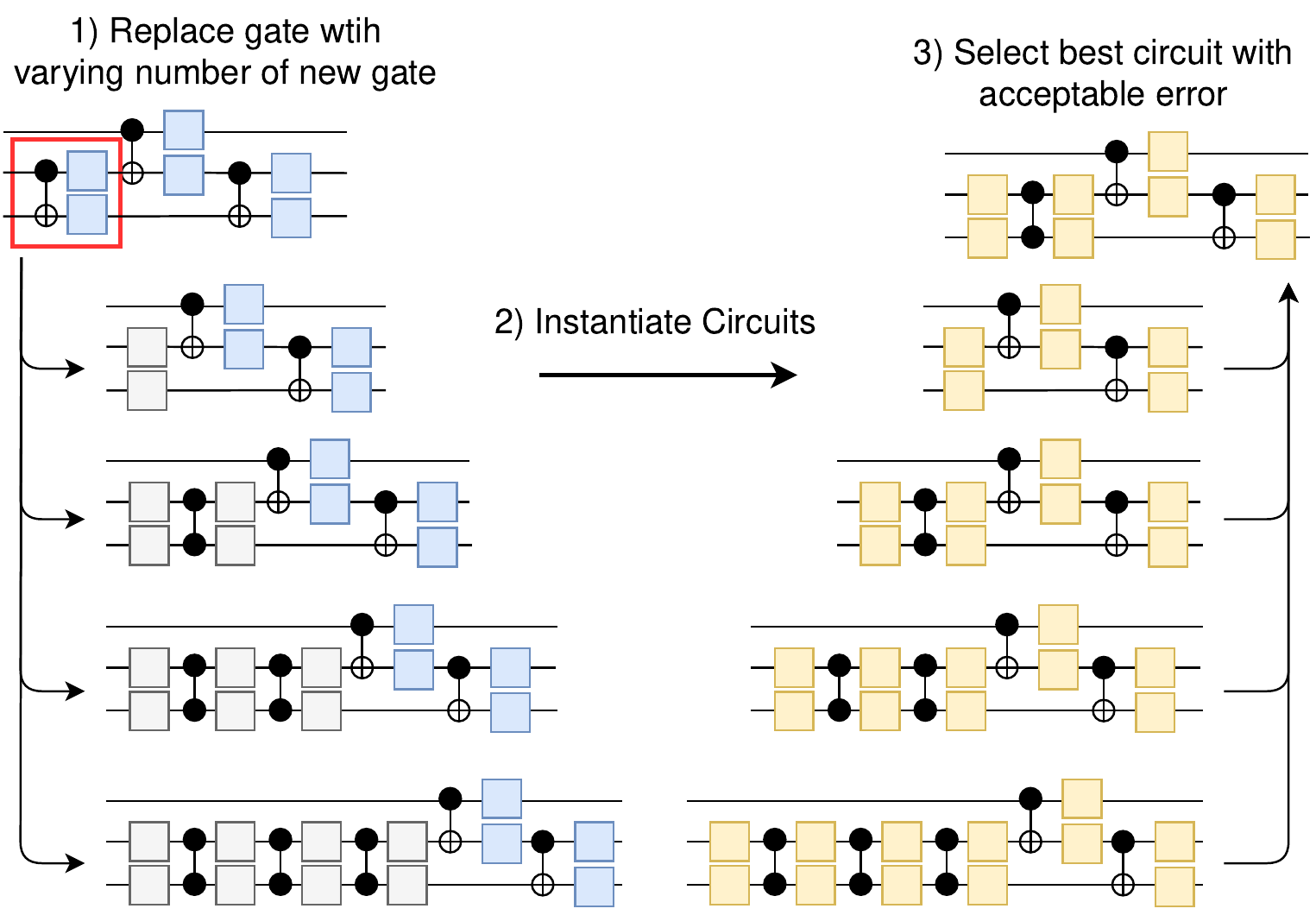}
    \caption{\footnotesize \it
        Instantiation can be used to convert gates from one type to another.
        This can be done by replacing the desired gate with a sequence of other gates
        and reinstantiating. Since this is not guaranteed to find a solution,
        we try a few different gate sequences, each one longer than the last.
        By default, we create 4 different sequences with 0, 1, 2, and 3 gates from the new gate-set respectively.
        After instantiating, we select the successful circuit with the fewest number of gates.
    }
    \label{fig:rebase}
\end{figure*}

    We propose a gate-set transpilation algorithm based on gate substitution
    using numerical instantiation. This works by iteratively replacing gates from
    one gate-set with gates from a target set. \comment{Figure~\ref{fig:rebase} provides an illustration of a single step of the algorithm.}
    Since single-qubit gates can always be grouped and decomposed into gates from any universal gate-set trivially,
    we focus on two-qubit gates in this algorithm.

  First, we group together consecutive two-qubit interactions (gates)
    that act on the same two qubits. These interactions will be
    replaced with gates in the new gate-set, one at a time in a
    left-to-right sweep over the circuit. Note that existing compilers
    use a similar strategy of forming a new generic two-qubit gate
    that is subsequently transpiled using either KAK decomposition or
    some other prescribed rule-based transformations. When introducing
    new two-qubit gates it may be non trivial to derive rules for
    transpiling from one source gate to the newly introduced one.
    
    Once a generic two-qubit gate is formed, our approach is based on
    the results from Bremner et al.~\cite{Bremner_2002} which shows
    that any two-qubit unitary can be represented by a short sequence
    of single qubit gates and arbitrary two-qubit entangling gates:
    the upper bound  on two-qubit entangling gates required can be easily
    calculated based on the single qubit gate set. This is the first main
    intuition behind our approach. The second intuition is that we can
    search over the space of these representations during
    instantiation. 

\begin{figure*}
        \centering
        \includegraphics{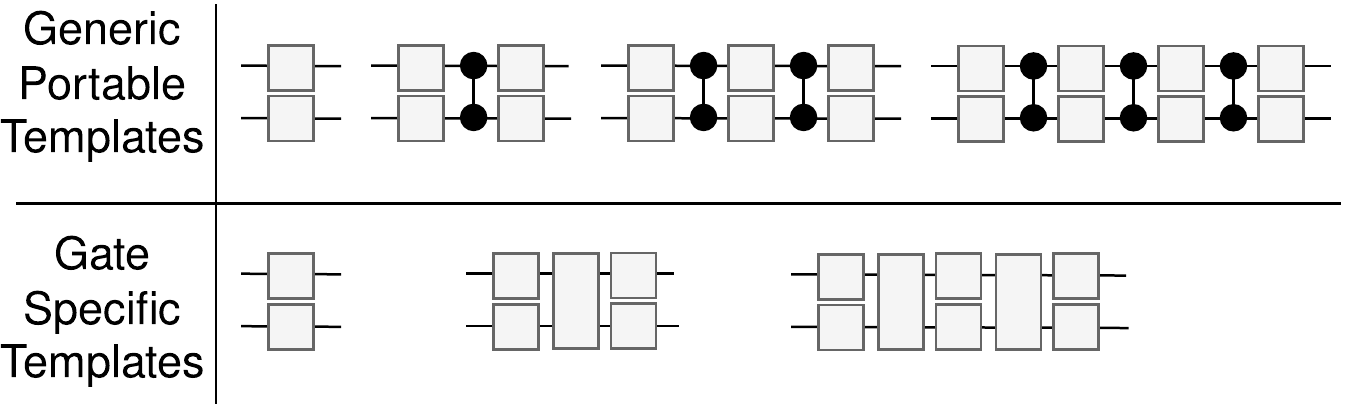}
        \caption{\footnotesize \it When using our retargeting algorithm, each gate from the old gate-set is replaced with templates containing new gate types. In the generic case, we use four templates with zero to three two-qubit gates. However, if we don't need all four templates, as in the case with some gates, we can reduce the number of candidate solutions instantiated improving runtime.}
        \label{fig:temps}
    \end{figure*}

    Thus, we attempt to replace each generic interaction found at the
    first step with a series of  templates created using gates in
    the new target gate-set. This collection is an enumeration of
    circuits starting with depth zero and ranging to an upper depth
    bound determined
    using~\cite{Bremner_2002}. \comment{Figure~\ref{fig:rebase} and
    }Figure~\ref{fig:temps} illustrate four templates with zero to
    three two-qubit gates in its place.  If a generic two qubit
    unitary can be implemented with fewer than N (three) gates in the
    new gate-set, this information can be used to reduce the number of
    candidate solutions instantiated, see Figure~\ref{fig:temps}. The
    results in this paper are obtained with the four shown templates. 
    
    The retargeting procedure is simple. We take the circuit in
    its original gate set and start traversing it, e.g. left to
    right. At each step, we replace a grouped two-qubit interaction from the circuit
    with all of the possible templates and reinstantiate. At each step
    of the algorithm, we have a collection of circuits equal to the
    number of template possibilities. Each of these circuits is
    composed of a prefix in the new gate set and a suffix in the
    original gate set.
    After instantiation, we select the shortest circuit with an acceptable error and continue the circuit scan.
    This process is described in pseudocode in Algorithm~\ref{alg:rebase_code} and illustrated in Figure~\ref{fig:rebase}.
    
    This retargeting algorithm is advantageous to rule-based or KAK-based retargeting
    algorithms for a few reasons.
    
    First, we consider more than the local two-qubit interaction when replacing gates.
    Both rule- and KAK-based methods consider isolated two-qubit interactions during their translation process, i.e. they freeze the rest of the circuit not allowing its parameters to change. By providing numerical instantiation with a global view of the circuit, we allow the entire circuit to morph as we plug in new gates. This leads to a greater ability to accommodate shorter sequences during translation.
    
    The second advantage is this algorithm's ease-of-use.
    Retargeting algorithms that rely on KAK decompositions
    or circuit identities require users to gather and input rules.
    Often when working with uncommon gate-sets, these rules can become
    difficult to produce. Numerical instantiation in this
    optimization algorithm removes the need to identify rules, enabling
    it to support any circuit without restriction. 
    
    Lastly, this algorithm can easily be adapted to target gate-sets with multiple
    two-qubit gates. To accomplish this, one can generate an additional set of candidate circuits for each additional two-qubit gate. Then after instantiation, the shortest acceptable circuit would be selected dynamically picking the gate providing the shortest implementation for each interaction.

\begin{algorithm}
\caption{Numerical Retargeting Algorithm}
\label{alg:rebase_code}
\begin{algorithmic}[1]
\State $C \gets \text{input circuit}$
\State $\epsilon_t \gets \text{acceptable error threshold}$
\State $U_t \gets \text{the unitary implemented by }C$
\State $G_i \gets \text{initial gate set}$
\State $G_t \gets \text{target gate set}$
\State $n \gets \text{max number of replaced gates per interaction}$
\While{$\exists g \in G_i \text{ s.t. } g \in C$} 
    \State Group g with neighbor gates
    \State $C_\text{candidates} = \{\}$
    \For{$j \in G_t$}
        \For{$i \in \{0, 1, \ldots, n\}$}
            \State $h \gets \text{block with }i\text{ gates of type } j$
            \State $C_c \gets C\text{ with }g\text{ replaced with }h$
            \State Push $C_c$ on $C_\text{candidates}$
        \EndFor
    \EndFor
    \State $C_\text{success} = \{\}$
    \For{$C_c \in C_\text{candidates}$}
        \State $C_c, \epsilon \gets \text{instantiate}(C_c, U_t)$
        \If{$\epsilon <= \epsilon_t$}
            \State Push $C_c$ on $C_\text{success}$
        \EndIf
    \EndFor
    \State $C \gets$ shortest circuit from $C_\text{success}$
\EndWhile
\end{algorithmic}
\end{algorithm}

\subsection{Partitioning and Scalability}
    
\begin{figure*}
        \centering
        
        \begin{subfigure}[t]{0.41\textwidth}
            \centering
            \includegraphics[width=\textwidth]{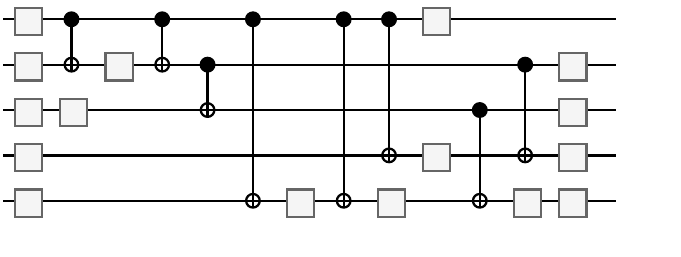}
            \caption{Original Circuit}
            \label{fig:unpart}
        \end{subfigure}
        ~
        \begin{subfigure}[t]{0.49\textwidth}
            \centering
            \includegraphics[width=\textwidth]{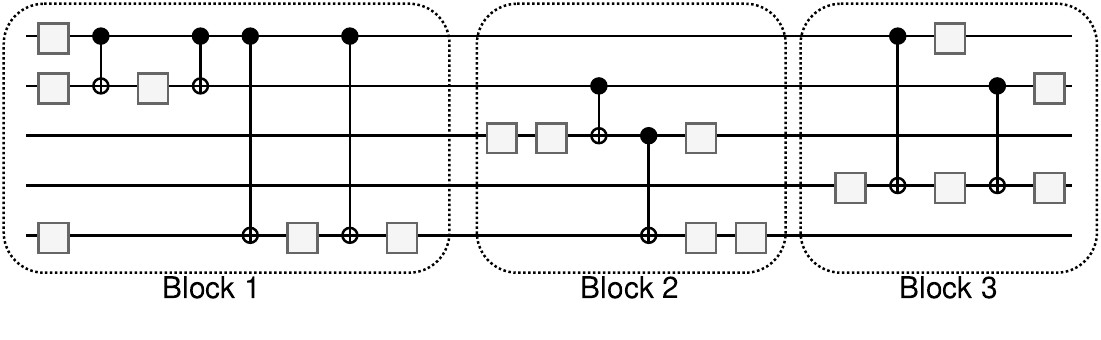}
            \caption{Partitioned Circuit with Three-Qubit Blocks}
            \label{fig:parted}
        \end{subfigure}
        
        \caption{\footnotesize \it The original circuit on the left is partitioned into blocks containing three qubits. This will group together consecutive gates acting on the same three-qubits. This allows algorithms making use of numerical instantiation to scale to very large circuits.}
        \label{fig:partitioning}
    \end{figure*}

    Since methodologies that use numerical instantiation scale
    exponentially with the size of the system being instantiated,
    we employ the circuit partitioning procedure described by Wu et
    al~\cite{qgo} to ensure competitive performance
    and scalability, see Figure~\ref{fig:partitioning}. For large
    circuits, we partition in smaller blocks which we can optimize
    (delete gates) or transpile directly, transform the blocks and
    reassemble the original circuit.
    
    There is a trade-off present with the size of partitions.
    One advantage of our optimization and translation algorithm
    is the ability to change non-local operations when instantiating
    Whenever we partition a circuit and restrict numerical instantiation
    within each block, we reduce the amount of non-local operations considered.
    Also, the partition size chosen determines the size of unitary matrices,
    and a larger matrix will lead to exponentially longer runtimes.
    With both algorithms, larger block sizes will likely lead to better quality
    results but smaller block sizes will complete quicker. In this
    work we evaluate forming  only blocks of three and four qubits.
\section{Experimental Setup}
\label{sec:exp}
    
    Both optimization and retargeting algorithms are implemented using the BQSKit framework~\cite{bqskit},
    and have been recently integrated into the main BQSKit package available at \texttt{https://github.com/bqskit/bqskit}.
    We evaluated them with Python 3.8.10 on a 64-core AMD Epyc 7702p
    system with 1TB of main memory.  We use the BQSKit provided
    instantiater configured to use the Google Ceres numerical
    optimizer with four initial starting points per instantiation. 
    We used the partitioning algorithm designed as part of QGo~\cite{qgo} and implemented in BQSKit configured to form 3-qubit blocks.
    
    We compared our optimization algorithm against three state-of-the-art optimizing compilers: Google's Cirq, IBM's Qiskit, and CQC's TKet. We also evaluated a workflow containing both our algorithm and these optimizing compilers. Our retargeting algorithm was compared against the same three tools, however, not all tools were able to target every gate we tested. For each gate-set, we only compared against compilers supporting it
    out-of-the-box. In all experiments we set an distance threshold of
    $\Delta = 10^{-10}$ for instantiation.

\subsection{Benchmarks}

    We used a benchmark suite consisting of real quantum algorithms of various different types and ranging in size from 5 to 64 qubits.
    There are multiplication and addition arithmetic circuits in the suite since they are commonly used to compare optimizing compilers \cite{ripple_adder, qgo}.
    Generated arithmetic circuits also tend to have ladders of two-qubit gates without single-qubit operations in between, which is valuable to test against. These circuits are known to be sub-optimal.
    Additionally, we have a couple of variational quantum algorithms
    in our benchmark suite. Two circuits using the Quantum Approximate
    Optimization Algorithm (QAOA)~\cite{qaoa} to solve the MaxCut
    problem with the hardware efficient
    ansatz~\cite{hardware_efficient}. Three circuits are simulating a spinful Hubbard model generated with the Bravyi-Kitaev mapping~\cite{bravyi_kitaev, hubbard}.
    We include a Grover circuit generated with Qiskit's algorithm library~\cite{grover}.
    Lastly, we include four time evolution circuits, two for
    Transverse Field Ising Models (TFIM)~\cite{tfimshin} and two for Transverse
    Field XY (TFXY) models. These circuits were generated by the constant-depth F3C++ compiler~\cite{constant_depth} and are believed to be the best known implementations. The benchmarks are listed in Figure~\ref{fig:og_nums} alongside their gate counts.
     
    \begin{figure}
    \centering
    \begin{tabular}{c|c|c}
        Benchmark & CNOT Gates & Single-qubit Gates  \\
        \hline
        adder9 & 98 & 64 \\
        adder63 & 1405 & 2885 \\
        mul10 & 163 & 107 \\
        mul60 & 11405 & 23666 \\
        qaoa5 & 42 & 27 \\
        qaoa10 & 85 & 40 \\
        hub4 & 180 & 155 \\
        hub8 & 2196 & 1513 \\
        hub12 & 8140 & 4932 \\
        grover5 & 48 & 80 \\
        tfim16 & 240 & 1200 \\
        tfim64 & 4032 & 20160 \\
        tfxy16 & 240 & 1200 \\
        tfxy64 & 4032 & 20160 \\
    \end{tabular}
    \caption{\footnotesize \it Gate counts for the 14 benchmarks before any optimizations or retargeting. The number of qubits in each circuit is included in the name as a suffix.}
    \label{fig:og_nums}
\end{figure}

\subsection{Verification}

\begin{figure}
    \centering
    \includegraphics{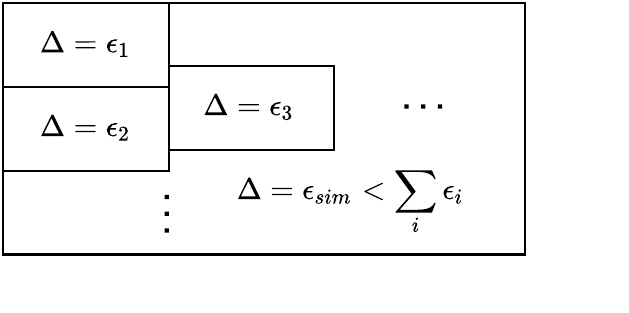}
    \caption{\footnotesize \it By partitioning large circuits in simulatable sections, we can calculate an upper-bound on error by summing together the individual section errors. As the section size increases, the upper-bound becomes more accurate.}
    \label{fig:verf}
\end{figure}

    We fully verified the results of our algorithms two ways.
    For circuits that were simulatable in size, we directly calculated
    the unitary error. All these circuits had a total error lower than $10^{-10}$, with a majority having zero error.
    For circuits that were unable to be directly simulated, we first partitioned the circuit into large simulatable sections. We then ran the algorithms within these sections, further partitioning to employ numerical instantiation as necessary. After the algorithms finished, we then directly computed the error of each section and calculated the theoretical upper-bound on process distance by summing them together~\cite{quest}, see Figure~\ref{fig:verf}.
    Section size matters when using this verification process. As the section size approaches the size of the circuit, the error upper-bound approaches the exact error of the circuit. For example with the 60-qubit multiplier circuit, there were 2150 three-qubit blocks. After running our optimization algorithm, summing the error on these small blocks gives an upper-bound of $9.97\times10^{-11}$; whereas, if we first partition the circuit into 8-qubit sections and use these to calculate the upper-bound, we get $2.61\times10^{-11}$. If we further increase the section size, it is likely the upper-bound will decrease more.
    
For all circuits, the total distance error $\Delta <
    10^{-10}$ when partitioning in three-qubit blocks. As simulating
    large circuits is computationally intensive, we have have
    performed the coarse grained procedure only for selected
    candidates to confirm the intuition that the error upper bound is
    indeed loose and the resulting circuits are accurate.

\section{Evaluation}
\label{sec:eval}

\subsection{Optimization Algorithm}
\begin{figure*}
    \centering
    
    \begin{subfigure}{\textwidth}
        \includegraphics[width=0.5\textwidth]{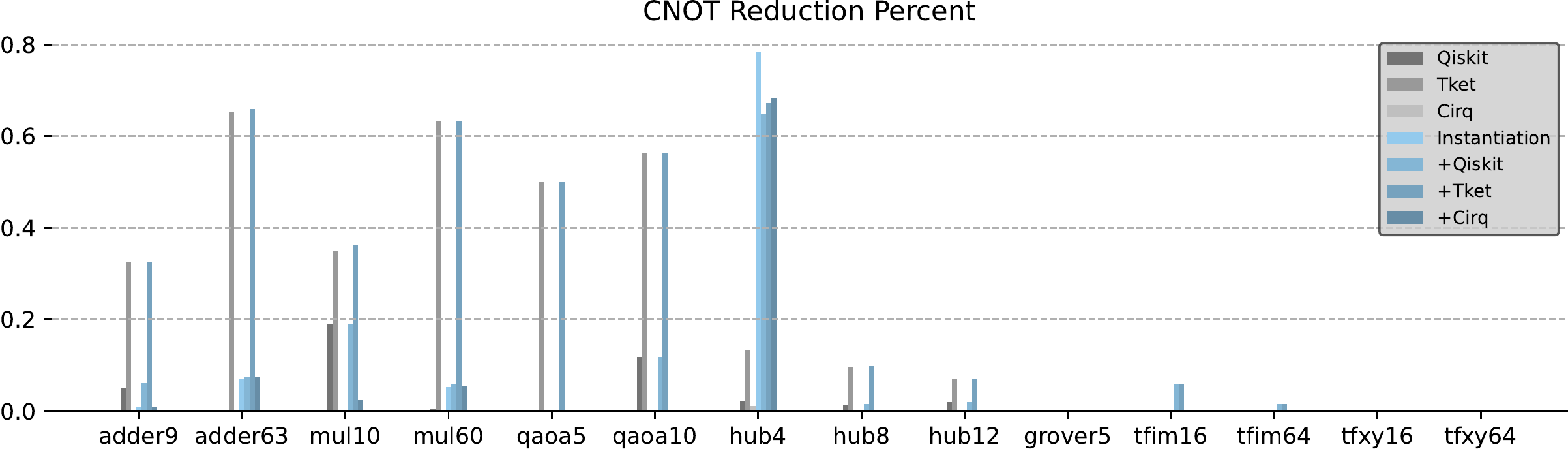}
        \includegraphics[width=0.5\textwidth]{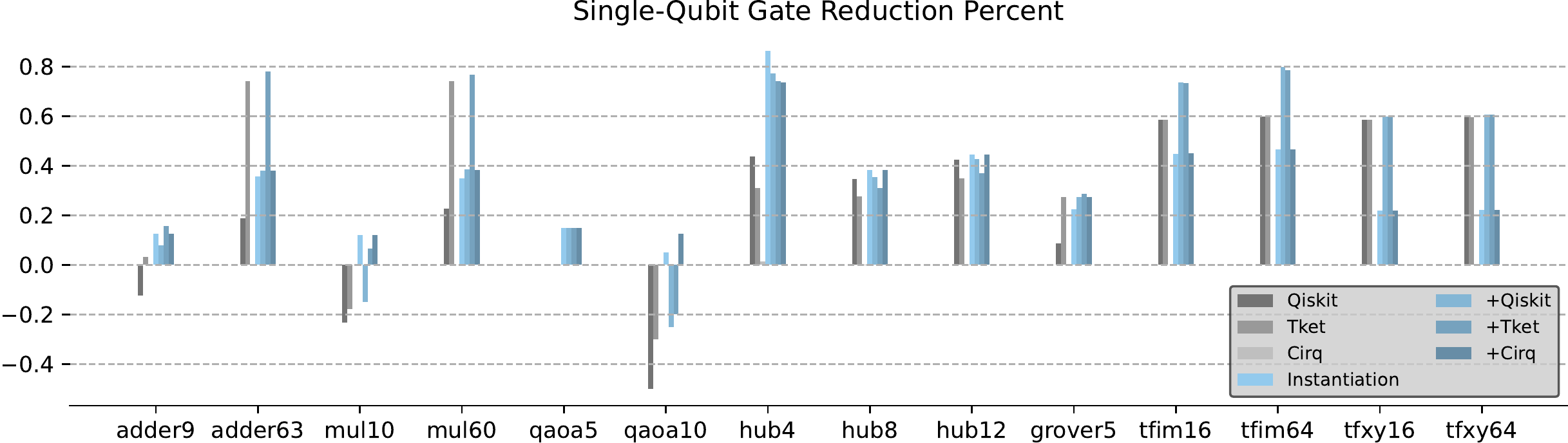}
    \end{subfigure}
    \par\bigskip
    
    \begin{subfigure}{\textwidth}
        \includegraphics[width=0.5\textwidth]{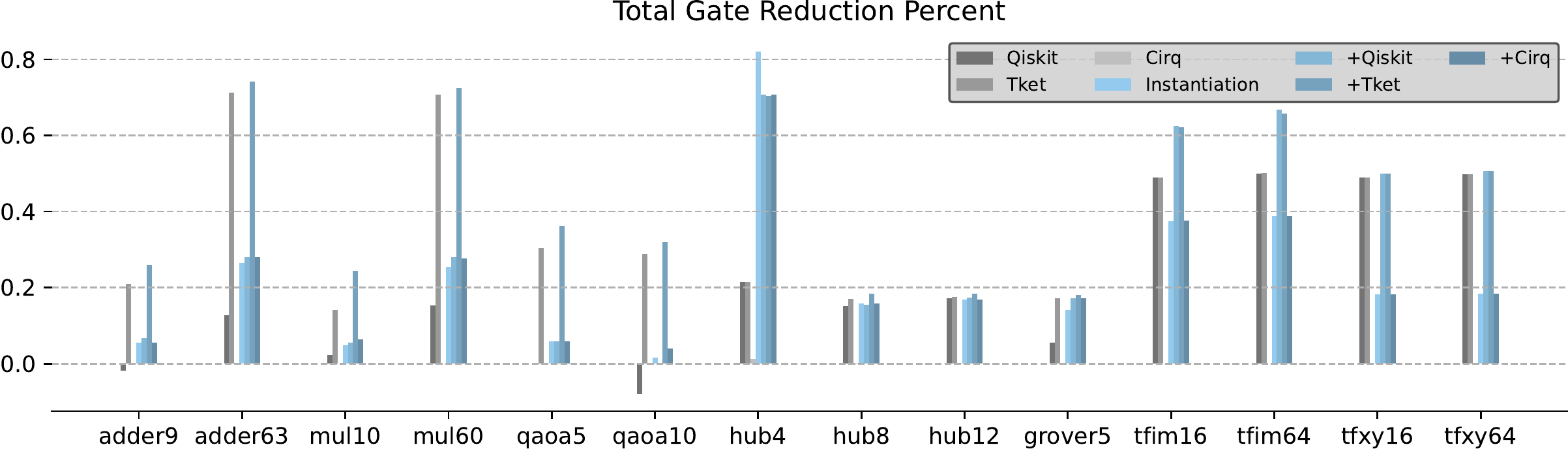}
        \includegraphics[width=0.5\textwidth]{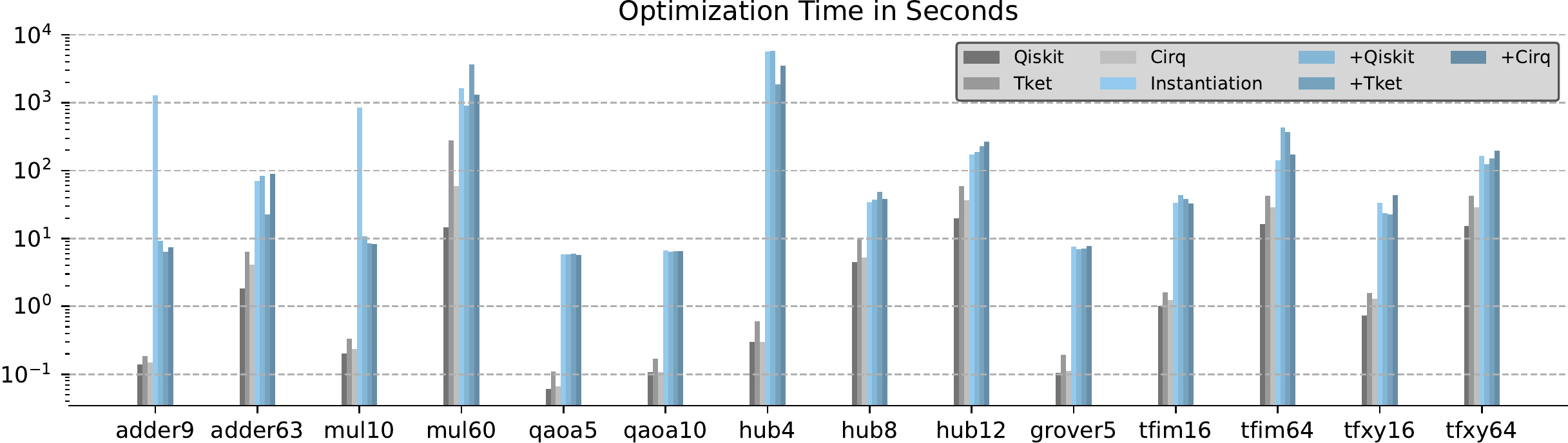}
    \end{subfigure}
    \par\bigskip
    
    \caption{\footnotesize \it The number of gates removed by different optimization algorithms as well as their runtime across a variety of different benchmarks. The number of gates removed are broken down into single-qubit and two-qubit gates removed. Additionally, the total number of gates removed are plotted. In some cases, more gates were added during optimization than removed, which is because some optimizers will prioritize two-qubit gates over single-qubit gates.}
    \label{fig:opt_graphs}
\end{figure*}

    The optimization algorithm is directly compared against Qiskit, Cirq, and Tket.
    Additionally, we include data from optimizing circuits after running them with the aforementioned tools, which are labeled ``+Qiskit'', ``+Cirq'', and ``+Tket''.
    The execution times and percent of gates reduced are plotted in Figure~\ref{fig:opt_graphs}.
    When run stand-alone and compared against compilers, our algorithm is able to improve on single-qubit gate count for 50\% of inputs, and on two-qubit gate count on 8\% of the inputs. When run after other tools, the optimizer produces shorter circuits in all cases, removing on average another 5\% of the two-qubit gates. When run together with either Qiskit or Tket, there was a reduction in two-qubit gate counts discovered in the TFIM circuits that was not possible with any individual optimizer.

    This behavior is explained by the interaction between original input circuit depth and partitioning: longer inputs form more partitions, with some loss of optimization potential at boundaries. Partition size further affects the quality of the optimization: larger partitions provide better opportunity for optimization at the expense of execution time.
    
    For example, the 4-qubit Hubbard model was the only circuit directly optimized without partitioning. This led to a 78\% reduction in CNOTs and a 86\% reduction in single-qubit gates, but required the longest compile time at 5728.9 seconds. The best result from the other compilers was only a 13\% reduction in CNOT counts for the same circuit. This illustrates  the trade-off with partitioning for scalability. By using larger blocks in partitioning or directly optimizing the circuit, we allow instantiation to consider more non-local interactions leading to a better optimization at the cost of runtime.
    
\subsection{Retargeting Algorithm}

\begin{figure*}
    \centering
    
    \begin{subfigure}{\textwidth}
        \includegraphics[width=0.5\textwidth]{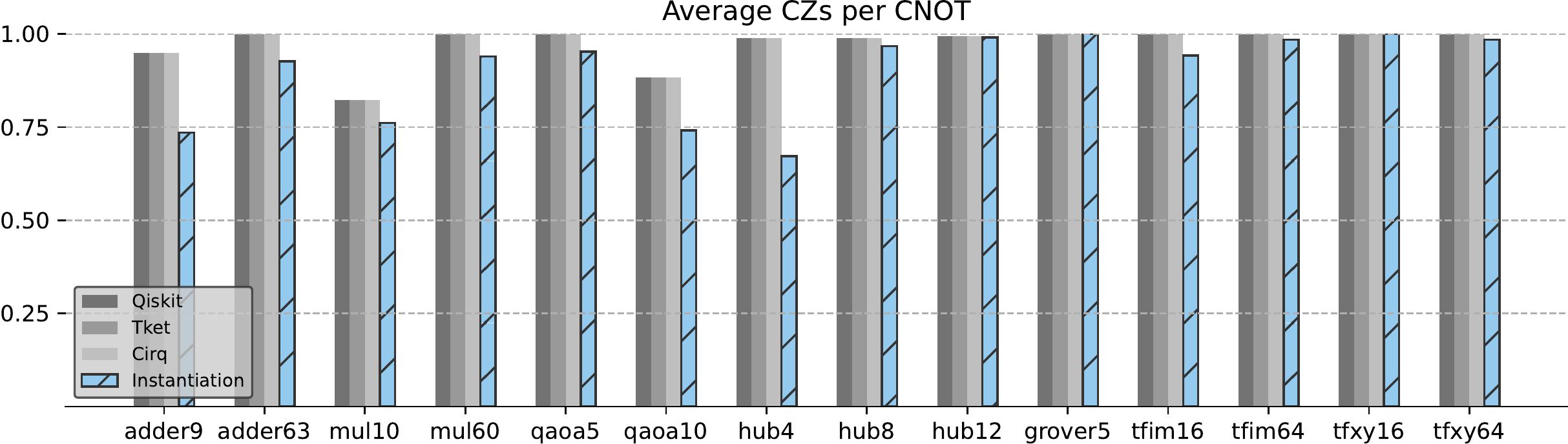}
        \includegraphics[width=0.5\textwidth]{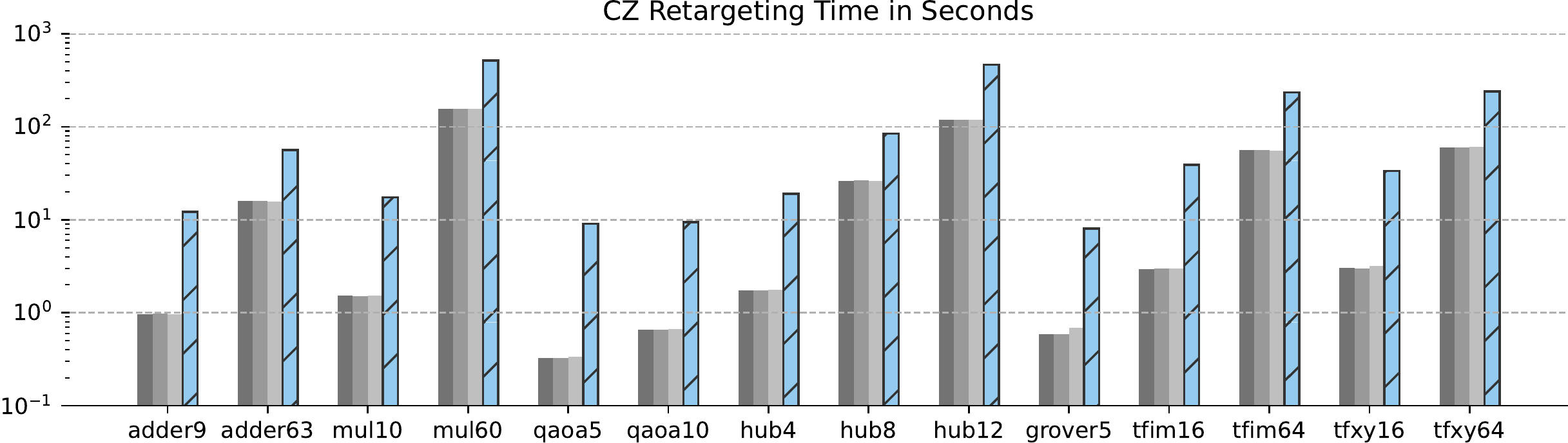}
    \end{subfigure}
    \par\bigskip
    
    \begin{subfigure}{\textwidth}
        \includegraphics[width=0.5\textwidth]{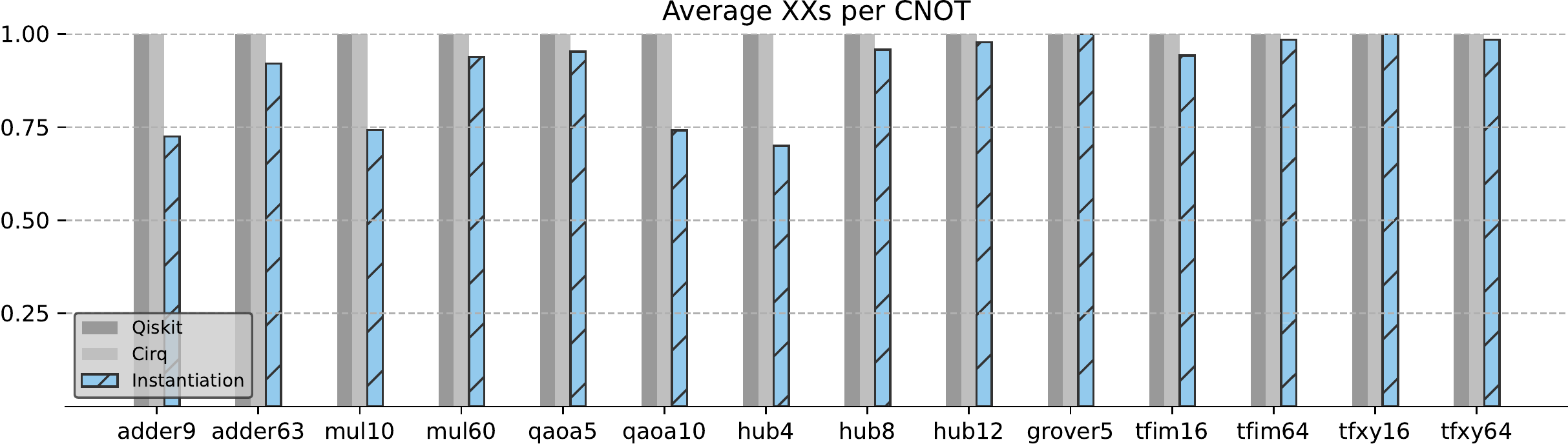}
        \includegraphics[width=0.5\textwidth]{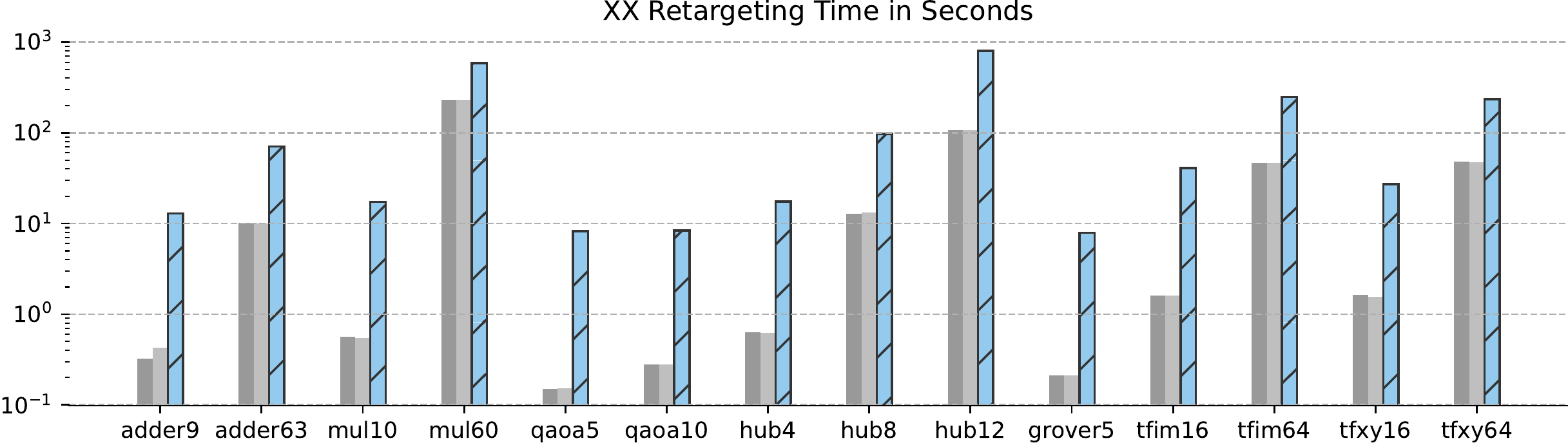}
    \end{subfigure}
    \par\bigskip
    
    \begin{subfigure}{\textwidth}
        \includegraphics[width=0.5\textwidth]{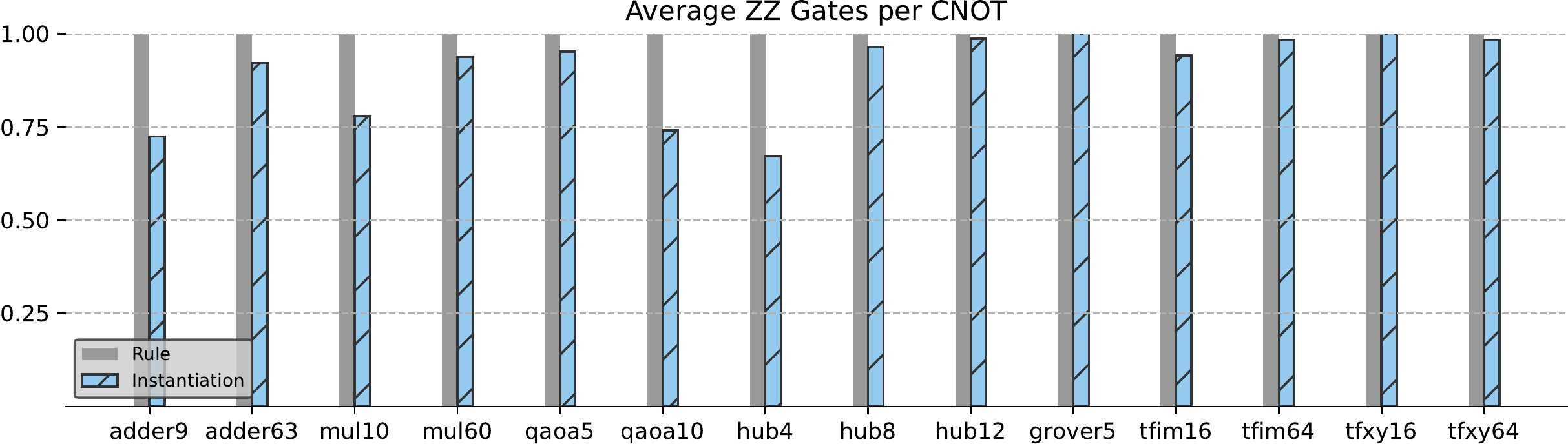}
        \includegraphics[width=0.5\textwidth]{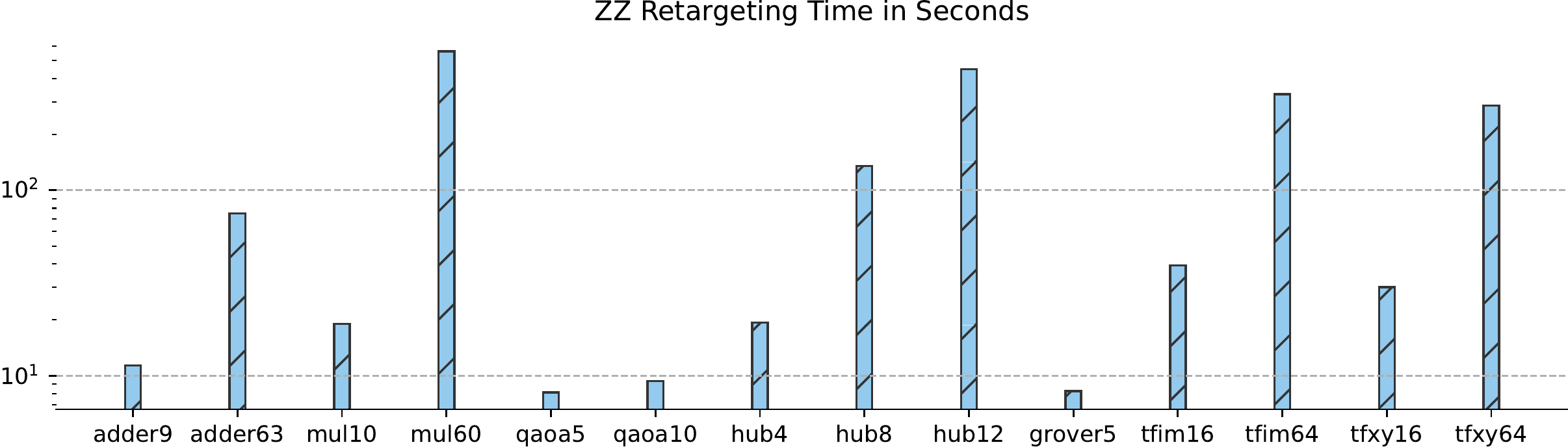}
    \end{subfigure}
    \par\bigskip
    
    \begin{subfigure}{\textwidth}
        \includegraphics[width=0.5\textwidth]{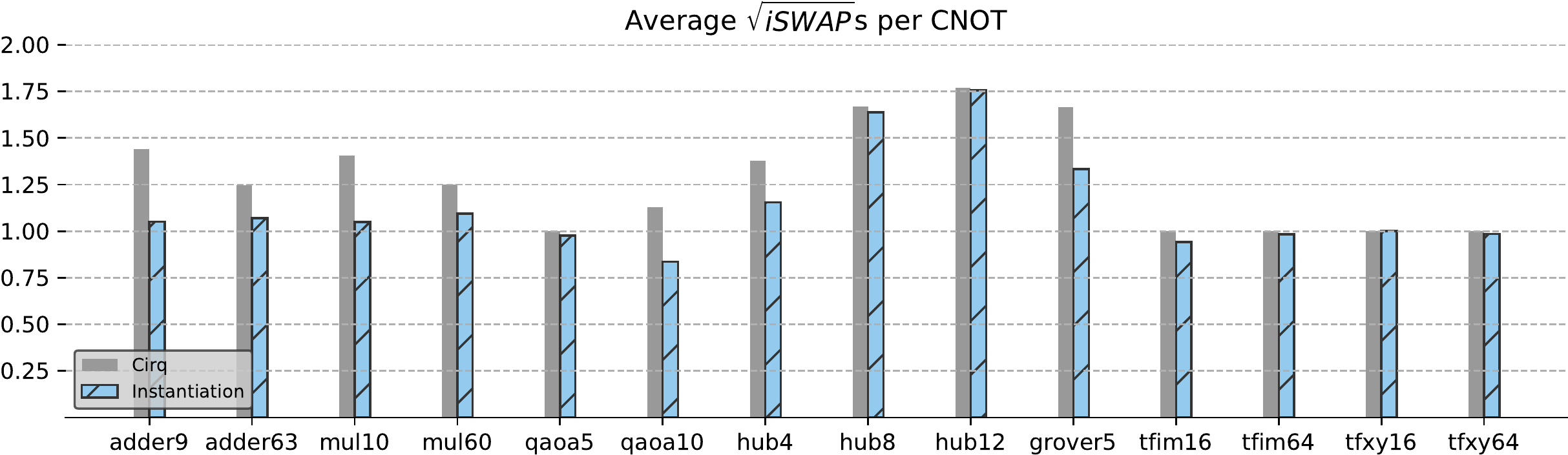}
        \includegraphics[width=0.5\textwidth]{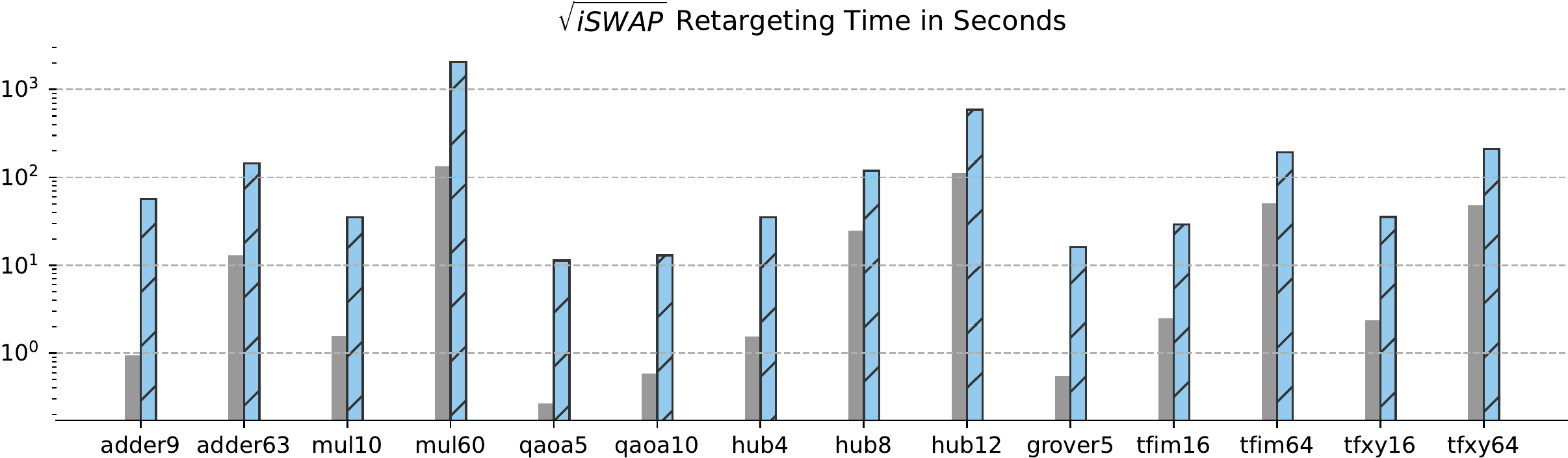}
    \end{subfigure}
    \par\bigskip
    
    \begin{subfigure}{\textwidth}
        \includegraphics[width=0.5\textwidth]{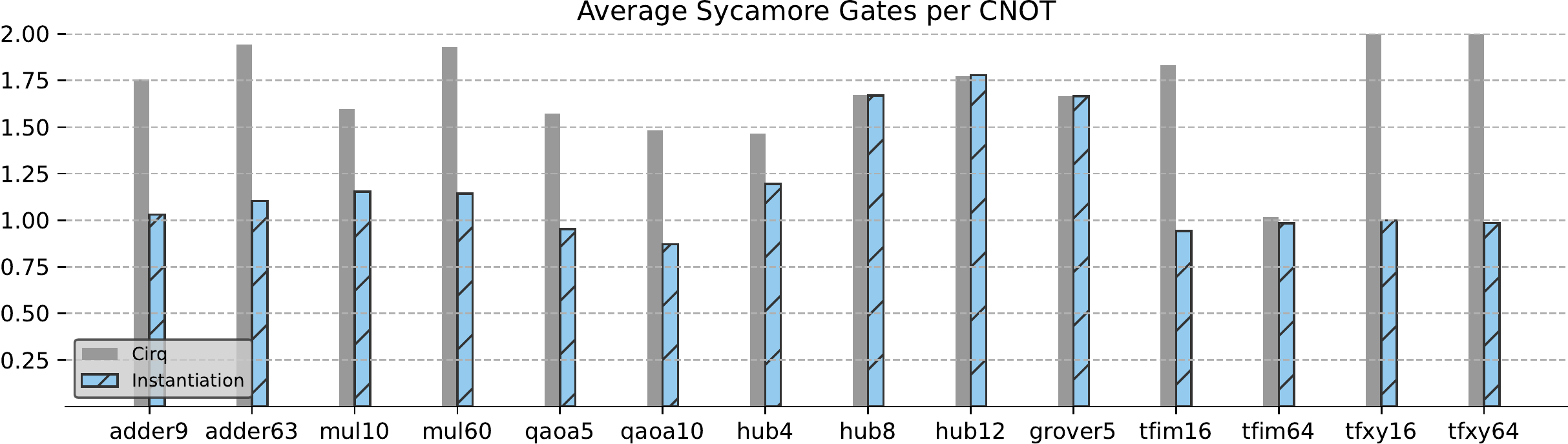}
        \includegraphics[width=0.5\textwidth]{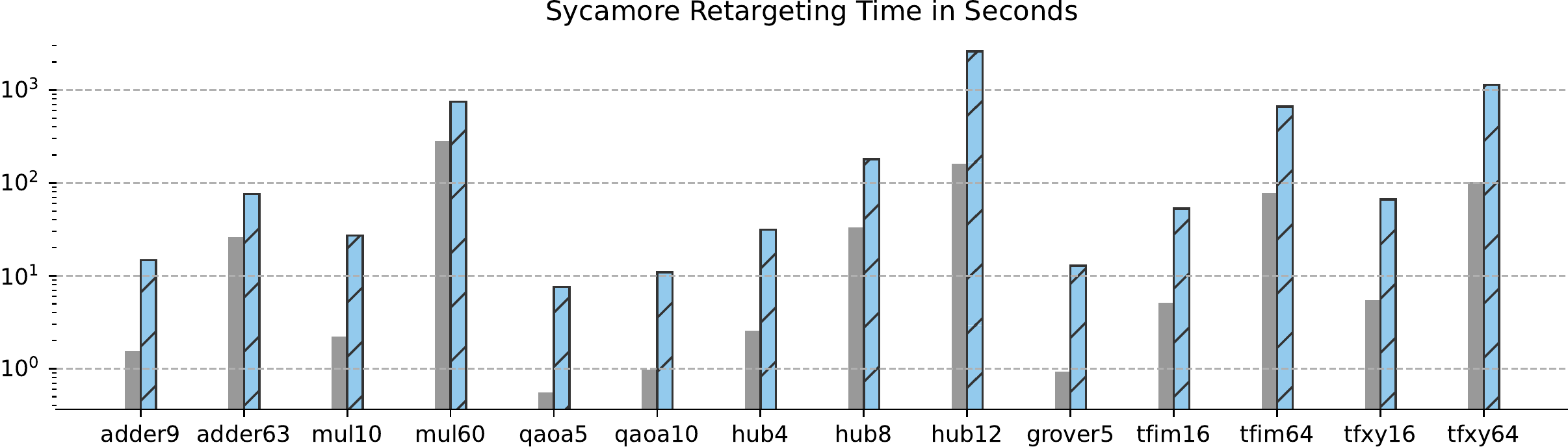}
    \end{subfigure}
    \par\bigskip
    
    \caption{\footnotesize \it Retargeting algorithm efficiencies and runtime across a variety different benchmarks. On the left, the ratio of final two-qubit gates to initial two-qubit gates is plotted for 5 different gate sets found in existing hardware. On the right, execution times for the different retargeting algorithms are plotted.}
    \label{fig:rebase_graphs}
\end{figure*}

    The retargeting algorithm was evaluated by converting the benchmark circuits to gate-sets that exist in hardware today: AQT and Rigetti's CZ gate~\cite{aqt_gate, rigetti_gate}, Honeywell's ZZ gate~\cite{honeywell_hardware}, IonQ XX gate~\cite{msgate}, and Google's Sycamore and \iswap~\cite{quantumsupremacy} gate. For each gate-set, we compared directly to compilers that support that gate-set. Neither Qiskit, Cirq, nor TKet support ZZ gate decomposition out-of-the-box, so we compared to the theoretical rule that translates 1 CNOT to 1 ZZ gate. The ratio of final two-qubit gate counts to initial two-qubit gate counts is plotted alongside execution times for the different benchmarks and compilers in Figure~\ref{fig:rebase_graphs}.
    
    The numerical instantiation based retargeting algorithm was able to translate circuits into new gate-sets using fewer two-qubit gates in all but one case. Google's Cirq was able to produce the 12-qubit Hubbard circuit in Sycamore Gates with 0.35\% fewer gates than this algorithm.
    
    When converting to CZ gates, all the other compiler produced roughly the same quality of result in roughly the same time. Our algorithm produced circuits with an average of 8\% fewer CZ gates and up to 32\% fewer CZ gates on the four-qubit Hubbard circuit.
    
    Qiskit and Cirq used a one-to-one rule to convert CNOT gates to XX gates, but did not optimize much past that. When compared our algorithm produced gates with 10\% fewer XX gates on average and a maximum of 30\% fewer gates for the four-qubit Hubbard circuit.
    
    In most cases we were able to beat the one-to-one rule for ZZ gates significantly with an average of improvement of 10\%. In the case of the four-qubit Hubbard model, we produced a result with 32\% fewer ZZ gates in the translated circuit versus CNOT gates in the input circuit.
    
    Converting to \iswap\  and Sycamore gates was only supported by Cirq, and these do not have a one-to-one rule to convert between CNOTs. Our algorithm produce circuits with 11\% fewer \iswap\  gates and 29\% fewer Sycamore gates. In some cases the difference between Cirq and our rebasing algorithm was quite significant. For example, the 64-qubit TFXY circuit compiled with our tool produced a circuit with 3970 sycamore gates, whereas Cirq produced a circuit with 8064 sycamore gates.
    
    Our algorithms relied on numerical instantiation rather than analytic methods, and as a result, we experienced a $14\times$ slowdown on average. The maximum slowdown for retargeting was compiling the 9-qubit adder to the \iswap\ gate which took a total of 56.46 seconds. The longest partitioned optimization execution was 1650.61 seconds for the 60-qubit multiply circuit.

\begin{figure*}
    \centering
    \includegraphics[width=\textwidth]{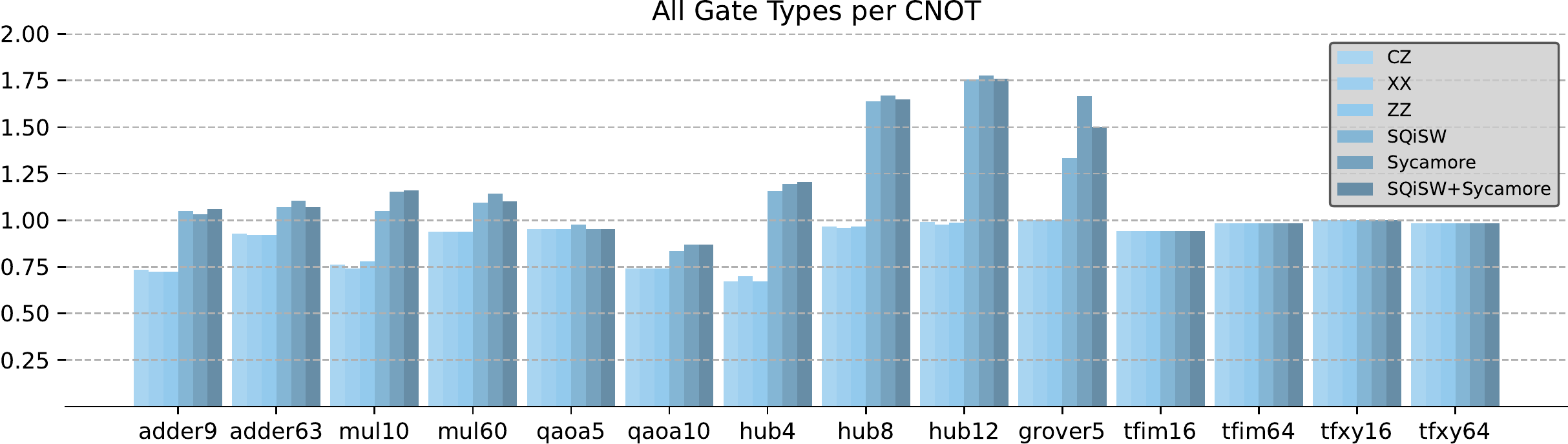}
    \caption{A comparison between all the gate-sets for each circuit produced with our instantiation-based algorithm.}
    \label{fig:all_ratios}
\end{figure*}

\subsection{Comparison of Gate-sets}

    The transpiled circuits are informative about the efficacy of different gate-sets
    across the algorithms. In Figure~\ref{fig:all_ratios}, we plot the final gate count as ratios to the original gate counts for all gate-sets. We also include the benchmarks compiled to a gate-set including both \iswap\  and the Sycamore gate.
    
    In every case, circuits including the \iswap\  or Sycamore gates required more gates than with the others. However, not all gates are equal. The Sycamore gate is executed in 12ns on Google's hardware, where the CZ gate requires 26ns~\cite{google_hardware}. In all cases, circuits in the Sycamore gate set required less than twice the number of CZs gates. This implies that, although the total circuit depth is greater in the Sycamore gate set, the total execution time and potentially decoherence may be less. This is especially true in the time evolution circuits, where the same amount gates were required regardless of the type.
    
    We also tried to test whether using more than one two-qubit gate is beneficial, and  compiled targeting  both  \iswap\  and Sycamore gates. The results are not conclusive, but it is probably premature to abandon this line of inquiry.  It may be the case that further tuning of our algorithm, or other approaches may improve on circuit quality.

    \subsection{Tuning the Algorithms}

    Throughout the evaluation, we ran our algorithms configured for four multi-starts and partitioned them into three-qubit blocks. While this is a recommended default and seems to produce good results, we can also trade circuit quality for execution speed: 1) bound the number of iterations during the optimization pass to improve execution speed; 2) configure for faster but ``inferior'' numerical optimization; and 3) increase circuit quality with bigger partitions/blocks at the expense of execution time.

The initial implementation of the {\tt mul60} circuit contains 11405 CNOT gates and 23666 single qubit gates.  With the default configuration, we optimized the 60-qubit multiplier circuit to 10798 CNOTs and 15384 single-qubit gates in 1650.61 seconds. When dialing the algorithm to improve performance at the cost of quality by using only one multi-start and only sweeping the circuit once during gate deletion we obtain 10865 CNOTs and 15765 single-qubit gates in only 119.82 seconds. This is about a $13\times$ speed-up at the cost of 0.6\% in quality.

	Similarly, we can configure the block width used during partitioning and the number of multi-starts for each instantiation when tuning the retargeting algorithm. With the same configuration used for other circuits, we produced a 60-qubit multiplier circuit with 12472 \iswap\ gates in 2045.45 seconds. If we increase the block width to four from three, we get 11746 \iswap\ gates in 6043.13 seconds. This configuration gives a 5.8\% improvement in quality with  $3\times$ execution  time increase.
\section{Discussion}
\label{sec:disc}
The retargeting algorithm proposed improves upon state-of-the-art techniques in both portability and quality, while the proposed optimization algorithm is competitive and augments well existing compilation workflows. Overall, the results are very encouraging for the adoption of numerical instantiation algorithms in compilers. We consider instantiation a good middle ground between ``peephole'' compilers and full-blown circuit synthesis: 1) it is more computationally intensive than compilers, but also able to improve circuit quality; 2) it has linear complexity when compared to the exponential complexity of full synthesis.

    We have shown how the algorithms can be tuned to execute in a time comparable to a traditional optimizing compiler. We note that further performance improvements are attainable through parallelization and exploiting GPU hardware. For the large circuits, partitioning results in an embarrassingly parallel problem easy to solve in distributed memory settings. 
    Current state-of-the-art numerical instantiaters rely on CPU-based optimization routines. There is a lot of active work efficiently simulating quantum circuits on GPUs~\cite{ cuquantum}, which could be valuable to getting instantiaters running on GPUs. Moving to GPUs may significantly speed up algorithms while also improving quality by batching more multi-starts. 

    The algorithms can also be adapted to process multiple gates or interactions in parallel. For example, instead of considering one two-qubit interaction at a time in the retargeting algorithm, we could attempt to change two or more interactions in every instantiation call. This change would reduce the total number of instantiation calls but potentially affect the quality.

    One practical question is where to place instantiation in a complete compilation workflow. Throughout our evaluations, we performed the algorithms on the reference circuits as they came and we never considered mapping or routing the initial inputs. Since both algorithms respect the logical connectivity of the input circuit, i.e., they never introduce a new gate on a pair of qubits not interacting in the input, they could be run post-routing. Furthermore, circuits with SWAP gates introduced during routing may benefit from our retargeting algorithm. Instead of applying the one-to-three SWAP gate conversion to CNOTs, which most compilers use, our retargeting algorithm can be run to reduce the ratio below three potentially. Our optimization algorithm can be run at multiple points in a standard compiler workflow. We can delete gates in the reference circuit, delete gates post-routing, and delete gates post-retargeting. Each time, new opportunities to remove gates may appear.

\section{Conclusions}
\label{sec:conc}

    In this work, we introduced and surveyed parameterized circuit instantiation and presented two numerical instantiation-based compiler algorithms for circuit optimization and gate-set transpilation. We demonstrated the tunability of our algorithms and verified our procedures using an error bound on very large circuits. Overall, these algorithms performed well and are a valuable addition to existing and new quantum compiler tools.

\section*{Acknowledgments}

This work was supported by the DOE under contract DE-5AC02-05CH11231,
through the Office of Advanced Scientific Computing Research (ASCR)
Quantum Algorithms Team and Accelerated Research in Quantum Computing programs.

\bibliographystyle{unsrt}
\bibliography{quantum, bibliography}

\begin{thebibliography}{10}

\bibitem{vqe}
Jarrod~R McClean, Jonathan Romero, Ryan Babbush, and Al{\'a}n Aspuru-Guzik.
\newblock The theory of variational hybrid quantum-classical algorithms.
\newblock {\em New Journal of Physics}, 18(2):023023, 2016.

\bibitem{qaoa}
Edward Farhi, Jeffrey Goldstone, and Sam Gutmann.
\newblock A quantum approximate optimization algorithm.
\newblock {\em arXiv preprint arXiv:1411.4028}, 2014.

\bibitem{qiskit}
Qiskit Developers.
\newblock {Qiskit: An Open-source Framework for Quantum Computing}, January
  2019.

\bibitem{cirq}
Cirq Developers.
\newblock Cirq, August 2021.
\newblock {See full list of authors on Github: https://github
  .com/quantumlib/Cirq/graphs/contributors}.

\bibitem{tket}
Seyon Sivarajah, Silas Dilkes, Alexander Cowtan, Will Simmons, Alec Edgington,
  and Ross Duncan.
\newblock t| ket>: a retargetable compiler for nisq devices.
\newblock {\em Quantum Science and Technology}, 6(1):014003, 2020.

\bibitem{bqskit}
Ed~Younis, Costin~C Iancu, Wim Lavrijsen, Marc Davis, Ethan Smith, et~al.
\newblock Berkeley quantum synthesis toolkit (bqskit) v1.
\newblock Technical report, Lawrence Berkeley National Lab.(LBNL), Berkeley, CA
  (United States), 2021.

\bibitem{qsearch}
Marc~G Davis, Ethan Smith, Ana Tudor, Koushik Sen, Irfan Siddiqi, and Costin
  Iancu.
\newblock Towards optimal topology aware quantum circuit synthesis.
\newblock In {\em 2020 IEEE International Conference on Quantum Computing and
  Engineering (QCE)}, pages 223--234. IEEE, 2020.

\bibitem{qfast}
Ed~Younis, Koushik Sen, Katherine Yelick, and Costin Iancu.
\newblock Qfast: Conflating search and numerical optimization for scalable
  quantum circuit synthesis.
\newblock In {\em 2021 IEEE International Conference on Quantum Computing and
  Engineering (QCE)}, pages 232--243. IEEE, 2021.

\bibitem{nacl}
Lukasz Cincio, Kenneth Rudinger, Mohan Sarovar, and Patrick~J Coles.
\newblock Machine learning of noise-resilient quantum circuits.
\newblock {\em PRX Quantum}, 2(1):010324, 2021.

\bibitem{squander1}
Péter Rakyta and Zoltán Zimborás.
\newblock Approaching the theoretical limit in quantum gate decomposition,
  2021.

\bibitem{squander2}
Péter Rakyta and Zoltán Zimborás.
\newblock Efficient quantum gate decomposition via adaptive circuit
  compression, 2022.

\bibitem{qgo}
Xin-Chuan Wu, Marc~Grau Davis, Frederic~T. Chong, and Costin Iancu.
\newblock Reoptimization of quantum circuits via hierarchical synthesis.
\newblock In {\em 2021 International Conference on Rebooting Computing (ICRC)},
  pages 35--46, 2021.

\bibitem{kak}
Robert~R Tucci.
\newblock An introduction to cartan's kak decomposition for qc programmers.
\newblock {\em arXiv preprint quant-ph/0507171}, 2005.

\bibitem{supermarq}
Teague Tomesh, Pranav Gokhale, Victory Omole, Gokul~Subramanian Ravi, Kaitlin~N
  Smith, Joshua Viszlai, Xin-Chuan Wu, Nikos Hardavellas, Margaret~R Martonosi,
  and Frederic~T Chong.
\newblock Supermarq: A scalable quantum benchmark suite.
\newblock {\em arXiv preprint arXiv:2202.11045}, 2022.

\bibitem{google_hardware}
Google devices.
\newblock \url{https://quantumai.google/cirq/google/devices}.
\newblock Accessed: 2022-04-29.

\bibitem{bestapprox}
Liam Madden and Andrea Simonetto.
\newblock Best approximate quantum compiling problems.
\newblock {\em ACM Transactions on Quantum Computing}, 3(2), mar 2022.

\bibitem{quest}
Tirthak Patel, Ed~Younis, Costin Iancu, Wibe de~Jong, and Devesh Tiwari.
\newblock Quest: systematically approximating quantum circuits for higher
  output fidelity.
\newblock In {\em Proceedings of the 27th ACM International Conference on
  Architectural Support for Programming Languages and Operating Systems}, pages
  514--528, 2022.

\bibitem{APOSMM}
Jeffrey Larson and Stefan~M. Wild.
\newblock Asynchronously parallel optimization solver for finding multiple
  minima.
\newblock {\em Mathematical Programming Computation}, 10(3), 2 2018.

\bibitem{Bremner_2002}
Michael~J. Bremner, Christopher~M. Dawson, Jennifer~L. Dodd, Alexei Gilchrist,
  Aram~W. Harrow, Duncan Mortimer, Michael~A. Nielsen, and Tobias~J. Osborne.
\newblock Practical scheme for quantum computation with any two-qubit
  entangling gate.
\newblock {\em Physical Review Letters}, 89(24), nov 2002.

\bibitem{ripple_adder}
Steven~A Cuccaro, Thomas~G Draper, Samuel~A Kutin, and David~Petrie Moulton.
\newblock A new quantum ripple-carry addition circuit.
\newblock {\em arXiv preprint quant-ph/0410184}, 2004.

\bibitem{hardware_efficient}
Abhinav Kandala, Antonio Mezzacapo, Kristan Temme, Maika Takita, Markus Brink,
  Jerry~M Chow, and Jay~M Gambetta.
\newblock Hardware-efficient variational quantum eigensolver for small
  molecules and quantum magnets.
\newblock {\em Nature}, 549(7671):242--246, 2017.

\bibitem{bravyi_kitaev}
Sergey~B Bravyi and Alexei~Yu Kitaev.
\newblock Fermionic quantum computation.
\newblock {\em Annals of Physics}, 298(1):210--226, 2002.

\bibitem{hubbard}
John Hubbard.
\newblock Electron correlations in narrow energy bands.
\newblock {\em Proceedings of the Royal Society of London. Series A.
  Mathematical and Physical Sciences}, 276(1365):238--257, 1963.

\bibitem{grover}
Lov~K Grover.
\newblock A fast quantum mechanical algorithm for database search.
\newblock In {\em Proceedings of the twenty-eighth annual ACM symposium on
  Theory of computing}, pages 212--219, 1996.

\bibitem{tfimshin}
Dongbin Shin, Hannes H{\"u}bener, Umberto De~Giovannini, Hosub Jin, Angel
  Rubio, and Noejung Park.
\newblock Phonon-driven spin-floquet magneto-valleytronics in mos\_2.
\newblock {\em Nature Communications}, 9(1):638, 2018.

\bibitem{constant_depth}
Lindsay Bassman, Roel Van~Beeumen, Ed~Younis, Ethan Smith, Costin Iancu, and
  Wibe~A de~Jong.
\newblock Constant-depth circuits for dynamic simulations of materials on
  quantum computers.
\newblock {\em Materials Theory}, 6(1):1--18, 2022.

\bibitem{aqt_gate}
Bradley~K Mitchell, Ravi~K Naik, Alexis Morvan, Akel Hashim, John~Mark
  Kreikebaum, Brian Marinelli, Wim Lavrijsen, Kasra Nowrouzi, David~I Santiago,
  and Irfan Siddiqi.
\newblock Hardware-efficient microwave-activated tunable coupling between
  superconducting qubits.
\newblock {\em Physical review letters}, 127(20):200502, 2021.

\bibitem{rigetti_gate}
Gates and instructions.
\newblock \url{https://pyquil-docs.rigetti.com/en/v2.7.0/apidocs/gates.html}.
\newblock Accessed: 2022-04-29.

\bibitem{honeywell_hardware}
Juan~M Pino, Jennifer~M Dreiling, Caroline Figgatt, John~P Gaebler, Steven~A
  Moses, CH~Baldwin, M~Foss-Feig, D~Hayes, K~Mayer, C~Ryan-Anderson, et~al.
\newblock Demonstration of the qccd trapped-ion quantum computer architecture.

\bibitem{msgate}
Klaus M\o{}lmer and Anders S\o{}rensen.
\newblock Multiparticle entanglement of hot trapped ions.
\newblock {\em Phys. Rev. Lett.}, 82:1835--1838, Mar 1999.

\bibitem{quantumsupremacy}
Frank Arute, Kunal Arya, Ryan Babbush, Dave Bacon, Joseph~C Bardin, Rami
  Barends, Rupak Biswas, Sergio Boixo, Fernando~GSL Brandao, David~A Buell,
  et~al.
\newblock Quantum supremacy using a programmable superconducting processor.
\newblock {\em Nature}, 574(7779):505--510, 2019.

\bibitem{cuquantum}
Sam Stanwyck, Harun Bayraktar, and Tim Costa.
\newblock cuquantum: Accelerating quantum circuit simulation on gpus.
\newblock {\em Bulletin of the American Physical Society}, 2022.

\end{thebibliography}

\end{document}